\documentclass{emulateapj}
\begin{document}

\title{The Distance to NGC 2264}
\author{Eric J. Baxter\altaffilmark{1}, Kevin R. Covey\altaffilmark{2}, August A. Muench\altaffilmark{2}, G\'{a}bor F\H{u}r\'{e}sz\altaffilmark{2}, Luisa Rebull\altaffilmark{3}, Andrew H. Szentgyorgyi\altaffilmark{2}}
\altaffiltext{1}{Astronomy and Astrophysics Department, University of Chicago, Chicago, IL 60637}
\altaffiltext{2}{Harvard Smithsonian Center for Astrophysics, 60 Garden St., Cambridge, MA 02138}
\altaffiltext{3}{Spitzer Science Center, M/S 220-6, 1200 E. California Blvd., Pasadena, CA 91125}

\keywords{methods: statistical --- stars: formation ---
  stars: rotation --- stars: distances --- open clusters: NGC 2264}

\begin{abstract}
We determine the distance to the open cluster NGC 2264 using a
statistical analysis of cluster member inclinations.  We derive
distance-dependent values of $\sin i$ (where $i$ is the inclination
angle) for 97 stars in NGC 2264 from the rotation periods,
luminosities, effective temperatures, and projected equatorial
rotation velocities, $v \sin i$, measured for these stars.  We have measured 96
of the $v \sin i$ values in our sample by analyzing high-resolution
spectra with a cross-correlation technique.  We model the observed
distribution of $\sin i$ for the cluster by assuming that member stars
have random axial orientations and by adopting prescriptions for the
measurement errors in our sample.  By adjusting the distance assumed
in the observed $\sin i$ distribution until it matches the modeled
distribution, we obtain a best-fit distance for the cluster.  We find
the data to be consistent with a distance to NGC 2264 of
$913$ pc.  Quantitative tests of our analysis reveals uncertainties of 40 and
110 pc due to sampling and systematic effects, respectively.  This
distance estimate suggests a revised age for the cluster of $\sim$1.5 Myrs,
although more detailed investigations of the full cluster membership are
required to draw strong conclusions.
\end{abstract}

\section{Introduction}

NGC 2264 is an open cluster in the Monoceros OB1 association containing a
large population of young stars.  The cluster has been the focus of a
number of studies of early stellar evolution dating back to the work
of \citet{Herbig}.  NGC 2264 is an ideal target for such studies
because of its low line-of-sight extinction and minimal optical
nebular emission \citep{Park}.  In addition, most background stars are
obscured from the cluster by the presence of a large molecular cloud
complex \citep{Herbig}.

Previous distance determinations for NGC 2264 have favored a distance
of 800 pc, although there has been considerable spread around this
value: 800 pc by \citet{Walker}, 760 pc by \citet{Sung}, 760-950 pc by
\citet{Flaccomio}, 750 pc by \citet{Mayne2008}; see \citet{Dahm} for
a complete summary of previous distance measurements.  
These distance estimates have been primarily derived by fitting empirically
or theoretically defined main sequences to the location of high mass
(B and A type) stars in the HR diagram.  Although extensive Stromgren
narrowband photometry is available in the literature \citep{Strom1971,Perez1989}, it has not yet been used to refine distances to these early type stars 
\citep[e.g., ][]{Anthony-Twarog1982}.

An improved distance estimate to NGC 2264 would lower the uncertainties in the
luminosities derived for cluster members, which in turn constrain the
cluster's age.  As one of the premiere laboratories for studying star
formation in the Milky Way, a better estimate of the age of NGC 2264
would further our understanding of processes relevant to early
stellar evolution such as angular momentum transfer and 
the lifetime of circumstellar disks.

In this paper, we determine the distance to NGC 2264 using a
statistical technique that relies on measured projected rotation
velocities, rotation periods, luminosities and effective
temperatures of the low mass K \& M type cluster members.  The technique was first developed by
\citet{Hendry} and has subsequently been used to find distances to the
Pleiades, the Taurus star forming region, and the Orion Nebula Cluster
~\citep{Odell, Preibisch, Jeffries}.  This method has the advantage of
being nearly independent of stellar evolutionary models.

\subsection{The Method}

In brief, we first measure the projected rotational velocities of
cluster members, $v \sin i$ (where $v$ is the tangential velocity of
the stellar surface at the equator and $i$ is the inclination of the
stellar rotational axis on the sky such that $i = 90^{\circ}$ implies an edge
on orientation and $i = 0^{\circ}$ implies a pole on orientation) from
existing high-resolution spectra of NGC 2264 members \citep{Furesz}.
An effective temperature, $T_{eff}$, is estimated for each star from
either its spectral type or dereddened photometry.  Luminosities, $L$,
are estimated for cluster members from measured magnitudes by assuming
a nominal value for the cluster distance, a prescription for the
cluster reddening, and a standard bolometric correction.  Stellar
radii are then calculated from the estimated luminosities and
effective temperatures using the Stefan-Boltzmann relation.  The final
data needed for the distance determination are rotation periods
obtained from fits to periodic variations in the stellar light curves.

Bringing all of these data together, we calculate
$\sin i$ for each star as
\begin{eqnarray}
\label{eq:sini}
\sin i = \frac{P \cdotp (v \sin i)}{2 \pi R_{D}},
\end{eqnarray}
where $P$ is the measured rotation period and $R_{D}$ is the stellar
radius (the subscript indicates that the measured radius is dependent
on the stellar luminosity, which depends on the assumed cluster
distance, $D$).

Eq. \ref{eq:sini} allows us to generate an observed distribution of
$\sin i$ that is dependent on the adopted cluster distance.  We then
model the distribution of $\sin i$ assuming that the rotational axes
of stars in the cluster are randomly oriented.  By scaling the input
cluster distance so that the observed $\sin i$ distribution matches
the predicted distribution, we obtain an estimate for the true cluster
distance.

We begin in $\S$\ref{data} with a description of the data used to
determine the distance to NGC 2264; $\S$\ref{vsinidetermination}
describes our measurement of projected rotation velocities by applying
a cross-correlation routine to high resolution spectra of cluster
members; $\S$\ref{distancedetermination} describes the distance
determination technique in detail; our results are discussed in
$\S$\ref{discussion}.

\section{Data}
\label{data}

As described above, our distance measurement relies on 
$\sin i$ values for stars in NGC 2264.  Four types of
data are needed to calculate $\sin i$ for an individual star using
Eq. \ref{eq:sini}: the star's period, luminosity, effective
temperature and projected equatorial rotational velocity, $v \sin i$.
Period ($\S$\ref{periodsub}), effective temperature
($\S$\ref{teffsub}) and luminosity ($\S$\ref{lumsub}) data were
obtained from a catalog compiled by \citet{Rebull} and are briefly described
below.  Most of the $v \sin i$ data, on the other hand, were
calculated from spectra ($\S$\ref{spectrasub}) using a cross-correlation
technique further discussed in $\S$\ref{vsinidetermination}.

\subsection{Periods}
\label{periodsub}

Rotation periods for pre-main sequence stars are determined by
measuring periodic variations in the objects' brightness.  These
variations arise from the presence of large ($\sim 40^{\circ}$
angular radius) starspots on the surfaces of these young, magnetically
active stars \citep{Herbst}.  The transit of the starspot(s) as the
star rotates diminishes the observed stellar luminosity on
the order of a few tenths of a magnitude.  This variation is quite
stable and can be used to derive precise periods.

\citet{Rebull} compiled periods measured for members of NGC 2264 by
\citet{Rebull_periods}, \citet{Makidon} and \citet{Lamm}.  The period
measurements range from roughly 0.5 to 29 days and are predicted to be
accurate to roughly $\delta P / P \approx 1\%$.

\subsection{Effective Temperatures}
\label{teffsub}

The catalog assembled by \citet{Rebull} includes $T_{eff}$ estimates
derived from low-resolution spectral types and from the stars' optical colors
using the $V-I$ vs. T$_{eff}$ relation presented by
\citet{Hillenbrand}.  Each star's $V-I$ color was dereddened prior to
this calculation as described by \citet{Rebull_periods}: stars with
spectra were dereddened so that their observed $R-I$ colors matched
the intrinsic colors of that spectral type on the zero age main
sequence (ZAMS) defined by \citet{Bessell1991}, \citet{Leggett1992}
and \citet{Leggett1998}.  The $V-I$ colors for stars without spectra were
dereddened assuming the modal reddening of members with measured
spectral types.  While the use of an overall reddening for the cluster
is less than ideal, out of a total of 97 stars for which we derive
$\sin i$ values, only 14 lack spectral types.  Thus, larger errors
associated with photometric based $T_{eff}$ measurements likely have a
negligible effect on our results.

T Tauri stars almost certainly do not have a single photospheric
temperature, however; observations of the weak T Tauri star V410 by
\citet{Herbst} indicated the presence of two large, polar star spots
with characteristic temperatures of 3100 K, in constrast to the star's
4400 K. photosphere.  Observations of larger ensembles of T Tauri
stars indicate similar photospheric--spot temperature differentials,
and typical spot covering fractions of $\sim$10\% \citep{Bouvier1989,
Johns-Krull2002}.

Somewhat counterintuitively, however, star--spots are unlikely to
introduce large errors in a star's derived photospheric temperature,
particularly for large spot--photosphere temperature differentials.
The star's integrated emission is dominated by non-spotted
photospheric flux, as the cooler spots emit significantly less flux
per unit area. Observational confirmation of this effect was provided
by \citet{Frasca2005}, who constructed detailed star--spot models of
RS CVn systems to reproduce broadband photometric light curves and
temperature sensitive line ratios from temporally resolved, high
resolution spectra of RS CVn systems.  While the best fit models
identified by \citet{Frasca2005} possessed photosphere--spot
temperature differentials of $\sim$1000 K, the temperature sensitive
spectroscopic line--ratios only departed from the photospheric value
by $\sim$150 K over the course of the observations.

In their Appendix A, \citet{Frasca2005} outline a formalism for
calculating the mean temperature observed from a star with a given
spot covering fraction and photosphere--spot temperature differential.
Using this formalism, we calculated the difference between a T Tauri
star's true (non-spotted) photospheric temperature and the temperature
that would be measured from its integrated (spot $+$ photsphere)
spectrum.  For typical T Tauri star parameters (T$_{phot}$=4000 K;
T$_{spot}$= 3200 K; f$_{spot}$ = 0.1), the temperature measured from
the combined spot$+$photosphere I--band spectrum would be 3992 K, only
8 K different from the `true' photospheric temperature.

Given the small size of this effect, we assume the uncertainties in
our derived temperatures are dominated by errors in the observed
colors, spectral types, and reddening corrections.  We expect the
temperature uncertainties introduced by these effects are roughly
$\delta T_{eff} / T_{eff} \approx 5 \%$.

\subsection{Luminosities}
\label{lumsub}

\citet{Rebull} calculated luminosities for cluster members from
dereddened photometry, assuming a distance to the cluster of 760 pc.
Stars without previously determined spectral types were dereddened by
the modal reddening value determined for stars with spectral types.
Not taking into account the uncertainty associated with the distance
assumption, the observed luminosities are likely accurate to within
$\delta L / L \approx 35 \%$.  This estimate includes contributions
from uncertainties in spectral type, extinction values, and
uncertainty due to source variability \citep{Hartmann_agespreads}.  

We note that the luminosity we calculate for each star from its
measured photospheric temperature will likely be a slight
over--estimate, as some fraction of the star's surface will be covered
with cooler, less luminous star--spots.  Again using the formalism
presented by \citet{Frasca2005} in their Appendix A, we calculated the
expected scale of this effect.  For our typical T Tauri star, the
bolometric luminosity calculated assuming a single measured
photospheric temperature (ie, T$_{obs}$=3992 K) is 5\% larger than the
true luminosity assuming the correct spot--photosphere differential
and covering fraction.  As the scale of this effect is well within our
errors, and as we lack a robust characterization of the spot properties
for the stars in our sample, we simply adopt the luminosity implied by
the single temperature associated with each star's spectral type.

\subsection{Echelle Spectra}
\label{spectrasub}

\begin{figure*}
\centering 
\includegraphics[scale = 0.5]{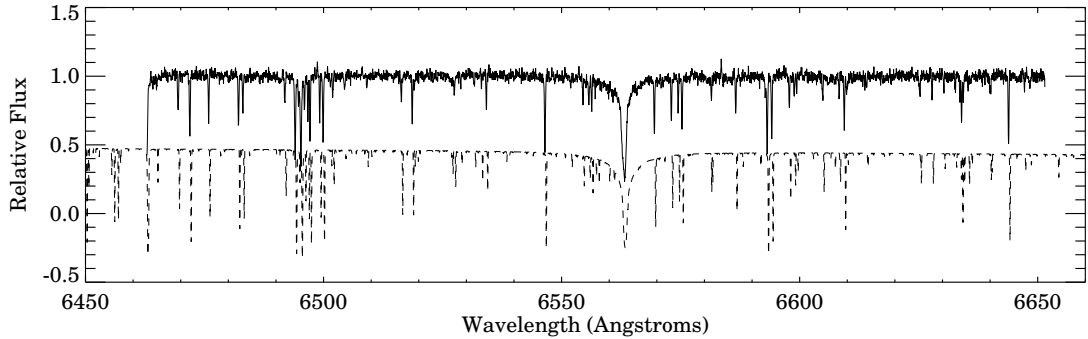}
\caption{A spectrum of a typical star in our sample (solid line).
 Also plotted (dashed line) is the best fitting template spectrum for
 this star (see $\S$\ref{Template_Grid}).  The template spectrum has
 been shifted to lower flux values for visual clarity.  As noted in
 the text ($\S$\ref{cc_parameters}) the H$\alpha$ region (from 6545
 \AA\,to 6585 \AA) was excluded from the cross-correlation procedure
 because of its strong variability between spectra.}
\label{typical_spectrum}
\end{figure*}

Spectra for 923 stars in and around NGC 2264 were obtained from the
sample observed by \citet{Furesz} using Hectochelle
\citep{Szentgyorgyi2006}, a multiobject echelle spectrograph located
on the 6.5 m Multiple Mirror Telescope (MMT).  The spectra cover a
wavelength range of 6450 \AA\,to 6650 \AA\, and have a resolution of
$R \sim 34,000$, providing velocity resolution of $\sim$ 9 km/s.  A
typical spectrum obtained for a star in NGC 2264 is shown in
Fig. \ref{typical_spectrum}.

\section{Projected Rotational Velocities}
\label{vsinidetermination}

\subsection{The Correlation Technique}

Because few stars in our initial catalog had previously measured $v
 \sin i$'s, it was necessary to extract this data from the
 spectra of \citet{Furesz}. To make this measurement, we used a
 cross-correlation technique similar to that developed by
 \citet{Tonry}. The correlation parameter, $C(v_{R})$, between a
 target (i.e. a cluster member) and a well-matched template spectrum
 is determined as a function of the radial velocity of the template
 spectrum, $v_{R}$.  $C(v_{R})$ is obtained by inverse Fourier
 transforming the product of the discrete Fourier transforms of the
 target and template spectra \citep{Hartmann}.  To generate
 $C(v_{R})$, we have developed a custom cross-correlation routine in
 IDL based on a heavily modified version of a routine originally
 developed by \citet{White}.

\begin{figure}[h]
\centering
 \includegraphics[scale = 0.55]{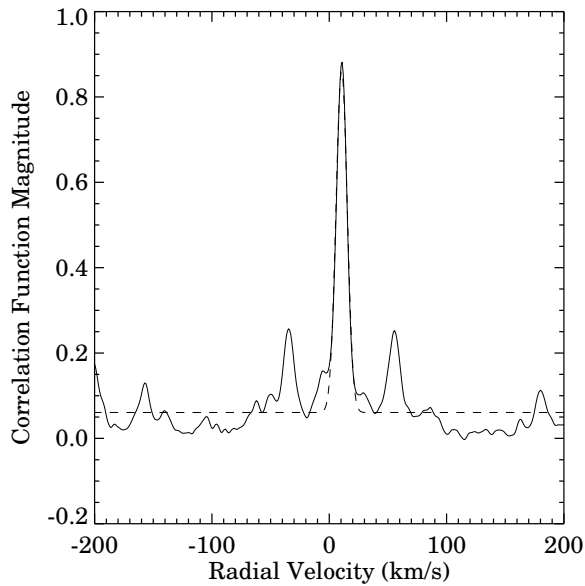}
\caption{A cross-correlation function resulting from the correlation
  of a target spectrum with the corresponding best fit template model.
  The cross-correlation function shows a clear peak at roughly 10 km/s;
  the width of the function formally implies a rotation velocity of about 3.4
  km/s, below our threshold for confident $v \sin i$ detections.  A Gaussian fit to the cross-correlation function is shown as
  a dashed line.  The above example represents a better than average
  cross-correlation result, with an $R$ value of approximately 44.}
\label{typical_cc}
\end{figure}

The location of the peak of $C(v_{R})$ corresponds to the value of
$v_{R}$ for which the target and template spectra are best matched.
Thus, the correlation procedure provides us with a measure of $v_{R}$
for stars in our sample.  Furthermore, because rotational broadening
of the target star's spectral lines results in a broader $C(v_{R})$
peak, the width of the $C(v_{R})$ peak, $\sigma_{C}$, provides a
measure of $v \sin i$ for the target star.  The correlation function
$C(v_{R})$ resulting from cross-correlating one of our target spectra
with a template spectrum is shown in Fig. \ref{typical_cc}.

The accuracy of the kinematic properties measured by the
cross-correlation technique depends on the quality of the agreement
between the strength and shape of the spectral features present in the
template and target spectra.  To ensure that the target and template
spectra are reasonably well matched, we correlate each target spectrum
against a grid of template spectra covering a range of temperatures
and surface gravities (see $\S$\ref{Template_Grid}).  The best
matching template is then used to derive $v \sin i$ for that target.
The degree of `matching' between a template spectrum and a target
spectrum is quantified with a statistical quantity, $R$, defined as
the ratio of the height of the maximum peak in $C(v_{R})$ to the
root-mean-square of the antisymmetric component of $C(v_{R})$
\citep{Tonry}.  In an idealized scenario in which the target spectrum
has no noise, the correlation function would be perfectly symmetric
around some particular value of $v_{R}$.  $R$ therefore measures the
strength of the correlation function peak against the noise in
$C(v_{R})$.  Thus, the template that produces the largest $R$ value
when correlated against a particular target can be said to be the
optimal template for that target.

\subsection{The Template Grid}
\label{Template_Grid}

We cross-correlated our targets against a set of templates calculated
 by \citet{Coelho}.  The template grid covers temperatures ranging
 from $3000$ K to $7000$ K in increments of $250$ K and $\log g$
 (surface gravity) values ranging from $0.0$ to $5.0$ in increments of
 $0.5$.  Because the strength of the H$\alpha$ line at 6562
 \AA\, exhibits strong dependence on accretion that could potentially
 interfere with our ability to derive $v \sin i$ values, each spectrum
 was divided into two regions bracketing H$\alpha$ (a `blue' region at
 wavelengths below 6545 \AA\, and a `red' region at wavelengths greater
 than 6585 \AA).  Both regions were then correlated separately with
 the template spectra.  Subsequent analysis revealed that our ability
 to derive accurate $v \sin i$ values was greater for the
 cross-correlations performed on the red region.  This effect was
 likely due to the fact that the $C(v_{R})$ peaks tended to have a
 more Gaussian shape for the red cross-correlations.  Thus, in future
 discussions $R$ refers to the value of $R$ calculated for the red
 region.

The relationship between $\sigma_{C}$ and the target $v \sin i$ was
calibrated separately for each template spectrum by cross-correlating
the template spectra with synthetic target spectra for which $v \sin
i$ is known.  The synthetic target spectra were created by introducing
artificial rotational broadening into the template spectrum and
degrading the spectral resolution and sampling to match the
Hectochelle data.  We performed this test for each template with
synthetic targets whose $v \sin i$'s ranged from 0 to 99 km/s in
steps of 3 km/s.  The best fitting template for the example target
spectrum in Fig. \ref{typical_spectrum} is also shown in that figure,
modified to reflect the corresponding $v_{R}$ and $v \sin i$ as
determined through our cross-correlation procedure.

\subsection{Parameters Derived from Cross-Correlation}
\label{cc_parameters}

\begin{figure}[h]
\centering \includegraphics[scale =0.5]{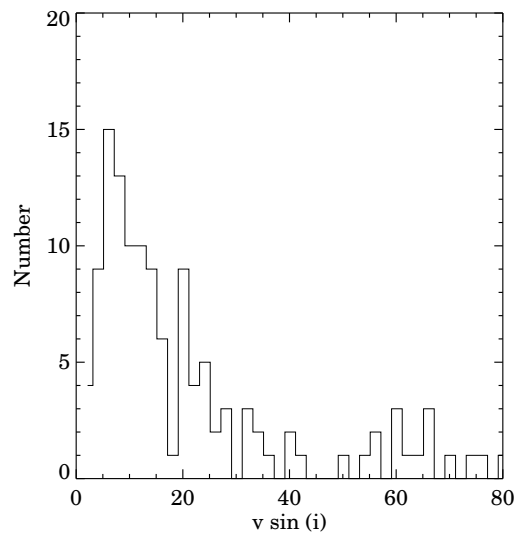}
\caption{The distribution of derived values of $v \sin i$.  As can be
  seen from the plot, the observed frequency of stars with a
  particular $v \sin i$ declines at $v \sin i \lesssim 6$ km/s as a
  result of an instrumental detection threshold as well as a true lack
  of stars with very low $v \sin i$.  }
\label{vsini_distribution}
\end{figure}

Once the best matching template spectrum for a target has been found,
the target's $v \sin i$ is measured by using the width of the
cross-correlation peak as an input for a fifth-order polynomial fit to
the $\sigma_{C}$-$v \sin i$ relation previously determined for that
template.  Valid $v \sin i$ values can only be derived when the
$\sigma_{C}$ for the target is within the range of $\sigma_{C}$
calculated for the template.  In total, valid $v \sin i$ values were
derived for 489 stars.  Fig. \ref{vsini_distribution} shows the
distribution of $v \sin i$ values for stars in our sample.  The
absence of stars with $v \sin i \lesssim$ 6 km/s likely represents a
combination of a true lack of stars with low $v \sin i$, as well as
the resolution limit of the Hectochelle spectrograph (9 km/s).  The finite
spectral resolution and standard measurement uncertainties produce
cross-correlation widths for low $v \sin i$ stars that are narrower
than the minimum $\sigma_{C}$ values in the template calibrations.
Our inability to separate the instrumental bias from the true
statistics at low $v \sin i$ motivated our decision to impose a cutoff
in $v \sin i$ (see $\S$\ref{sampleselection}).

\subsection{Uncertainties in $v \sin i$}
\label{testing_uncertainties}

Rigorous uncertainties were calculated for the $v \sin i$ values
derived from the cross-correlation procedure by running a Monte Carlo
test on a set of artificial spectra.  The synthetic spectra were
rotationally broadened to arbitrary $v \sin i$ and then matched to the
resolution and sampling of the Hectochelle data.  Gaussian noise was
also added to the synthetic targets to simulate the effects of a
finite signal to noise ratio in the actual data.  The 1-$\sigma$
amplitude of the noise was one fourth the mean signal level,
characteristic of the lowest quality data in the Hectochelle
observations.  Since the value of $v \sin i$ is known for the
synthetic spectra, the difference between the calculated $v \sin i$
and the true $v \sin i$ for the synthetic spectra provides a measure
of the uncertainties inherent in our cross correlation procedure.

\begin{figure}[h]
\centering
\includegraphics[scale = 0.5]{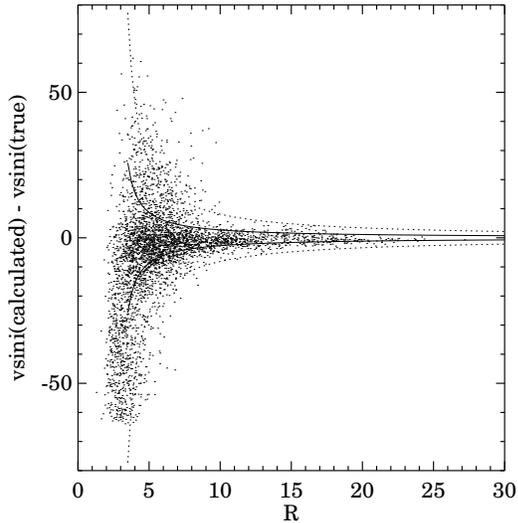}
\caption{Error in the $v \sin i$ values calculated by our
  cross-correlation routine as a function of $R$.  Each data point
  represents a synthetic target that has been rotationally broadened
  to some $v \sin i$(true) and then correlated against the template
  grid to find the best matching template.  $v \sin i$(calculated) was
  then determined from the relationship between $v \sin i$ and the
  width of the correlation function peak.  The solid line shows a fit 
  to the 1-$\sigma$ width of the distribution, while the dashed line
  represents a similar fit to the 3-$\sigma$ width of the
  distribution (see text for fit parameters).  }
\label{deltavsini}
\end{figure}

Uncertainties in $v \sin i$ were determined as a function of $R$ for
each synthetic target.  Fig. \ref{deltavsini} shows the deviation,
$\Delta v \sin i$, between the input and calculated value of $v \sin
i$ for each synthetic target as a function of $R$.  As should be
expected, $\Delta v \sin i$ decreases with increasing $R$ since higher
$R$ corresponds to a better match between the target and template.
Following \citet{Hartmann}, we quantify the $\Delta v \sin i$ vs. $R$
relation by fitting a curve of the form $a / (b + R)$ to the $1
\sigma$ width of $\Delta v \sin i$.  We find that the best fit is
$\Delta v \sin i = (19.7 \rm{\,\,km/s})/ (R - 2.74)$.  We apply this
relation to the R values measured from actual target spectra to
quantize our errors in measured values of $v \sin i$.

\section{Distance Determination}
\label{distancedetermination}

The combination of period, luminosity, effective temperature and $v
\sin i$ data allows us to calculate $\sin i$ for each of the stars in
our sample.  Since $\sin i$ depends on the absolute luminosities of
stars in our sample, the observed distribution of $\sin i$ values is
inherently dependent on the assumed cluster distance.  We use a Monte
Carlo routine to produce model $\sin i$ distributions that assume
randomly oriented stellar rotation axes and that incorporate the
effects of observational uncertainties.  The distance to NGC 2264 is
then constrained by comparing model $\sin i$ distributions to observed
distributions with different assumed cluster distances.  The distance
at which the modeled and observed distributions of $\sin i$ agree is
our best fit distance.

\subsection{Calculating $\sin i$}
\label{observed_sinis}

Assuming that the measured period, $P$, of each star is equal to its
rotation period at the equator, we have
\begin{eqnarray}
\label{siniR}
\sin i &=& \frac{P \cdotp (v \sin i )}{2 \pi R},
\end{eqnarray}
where $R$ is the stellar radius.  $P$ and $v \sin i$ are direct
observables, but $R$ must be inferred from each star's luminosity,
$L$, and effective temperature, $T_{eff}$, using the Stefan-Boltzmann
relation.  $L$ must itself be estimated from each star's
extinction-corrected bolometric magnitude ($m_{bol}$) and an adopted
distance to the cluster:
\begin{eqnarray}
\label{lmagdis}
L = 10^{-0.4(m_{bol} - 5 \log D + 5)}.
\end{eqnarray}
Thus, we can write an expression for $\sin i$ that is explicitly
dependent on distance:
\begin{eqnarray}
\label{sinidis}
\sin i = \frac{P (v \sin i) T_{eff}^{2} \sqrt{\sigma /
    \pi}}{10^{-0.2(m_{bol} - 5 \log D + 5)}}.
\end{eqnarray}
Since luminosities were previously calculated by \citet{Rebull} assuming a
distance $D_{0} = 760 $ pc, it is convenient to re-write Eq. \ref{sinidis} as 
\begin{eqnarray}
\label{siniL}
\sin i = \frac{P (v \sin i) T_{eff}^{2} \sqrt{\sigma / \pi}}{\sqrt{L_{0}}
  \left( D / D_{0} \right)},
\end{eqnarray}
where $L_{0}$ is the luminosity that has been calculated by
\citet{Rebull} for the stars in our sample.  Eq. \ref{siniL} allows us
to calculate the $\sin i$ distribution of stars in NGC 2264 from $P$,
$v \sin i$, $T_{eff}$ and $L_{0}$ data, as well as an initial estimate
of the cluster distance, $D_{0}$.

\subsection{Selection of Sample Stars}
\label{sampleselection}

Beginning with our sample of 923 stars for which we have Hectochelle
spectra, we use a series of quality cuts to identify those stars that
will allow us to produce an unbiased estimate of the distance to NGC
2264.  We first restrict our analysis to the 489 stars for which we
are able to measure $v \sin i$ using our cross-correlation routine.
We further restrict our sample to those stars with a
high-likelihood of being bona-fide members of NGC 2264.  Since the
cluster is a coherent kinematic system, imposing a radial velocity cut
( 10 km/s $< v_{R} <$ 30 km/s) \citep{Furesz} on our spectroscopic
sample identifies 273 likely members. Of these, 130 have the ancillary
measurements ($P$, $T_{eff}$, and $L_R$) necessary to compute $\sin
i$.

\citet{Rebull} identified a number of stars as likely members of NGC
2264 based on their positions in the sky and their location in
color-magnitude space.  We add one star to our $\sin i$ sample from
the catalog of NGC 2264 cluster members compiled by \citet{Rebull};
this additional star is the only catalog member without a Hectochelle
spectrum but with a previous $v \sin i$ measurement, as well as the
other measurements necessary to estimate $\sin i$.  Finally we
restrict our sample to the 97 stars with measured $v \sin i$ values
larger than 9 km/s, the minimum velocity resolution of our Hectochelle
data.  In the discussion that follows, we refer to this final subset
of 97 stars as the `distance sample'; for clarity we summarize the
steps in its selection in Table \ref{tab:sampleselection}.

\begin{deluxetable*}{lc}
\tablewidth{0pt}
\tabletypesize{\scriptsize}
\tablecaption{Summary of Distance Sample Selection \label{tab:sampleselection}}
\tablehead{
   \colhead{Sample Subset} &
   \colhead{Number of stars}}
\startdata
$\bullet$ Initial set of \citet{Furesz} spectra & 923 \\
$\bullet$ Reliable $v \sin i$ results & 489 \\
$\bullet$ Radial velocity members w/ $v \sin i$ & 273 \\
$\bullet$ Ancillary data for $\sin i$ estimate, & \\
 radial velocity member \& $v \sin i$ & 130 \\
$\bullet~v \sin i > 9$, ancillary data, & \\
 \& radial velocity member & 96 \citep[+1 from ][]{Rebull} \\ 
\enddata
\end{deluxetable*}

Table \ref{tab:sinidata} lists all the data used to calculate $\sin i$
for the 97 stars in the distance sample.  The `$L_{bol}$ \& $T_{eff}$
source' column identifies the type of reddening correction applied to
each star's photometry in estimating its luminosity and temperature;
stars with an `S' have individual reddening estimates based on
observed spectral types, while stars labeled `P' had photometry
corrected assuming the modal extinction derived for cluster members by
\citet{Rebull}.  Finally, the star added to the distance sample using
the $v \sin i$ measurement cataloged by \citet{Rebull} lacks a
corresponding $v \sin i$ error estimate: its entry in Table
\ref{tab:sinidata} lists `None' in the $v \sin i$ error column.

\begin{figure}[h]
\centering
\includegraphics[scale = 0.5]{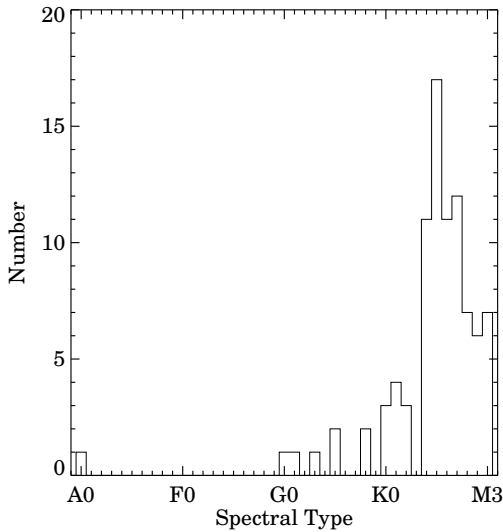}
\caption{The distribution of spectral types in the distance sample.
The bulk of the stars in our sample have types between K4 and M3.}
\label{spectypedist}
\end{figure}

Figure \ref{spectypedist} shows the distribution of spectral types for
stars in the distance sample.  Most of the stars in the distance
sample have spectral types in the range of K4 to M3.

\subsection{Error Distributions for Observed Parameters}
\label{error_distribution}

Observational uncertainties affect the shape of the measured $\sin i$
distribution.  In order to obtain a reliable distance estimate, we
must account for these uncertainties in our modeled values of $\sin
i$.  To do this, we first estimate uncertainties in $P$, $v \sin i$,
$T_{eff}$ and $L$ from the observational data; a Monte Carlo
simulation is then used to incorporate these error distributions into
the modeled value of $\sin i$.

The rotation periods of T Tauri stars can be measured with high
precision from their light curves.  The errors associated with such
measurements are usually on the order of 1\%.  In some cases,
confounding factors such as aliasing in the light curves or the
presence of multiple starspots can increase these errors dramatically
\citep[e.g.][]{Herbst02}.  However, the number of cases in which these
effects occur is typically small. Since the period errors are very
small compared to the errors associated with other T Tauri
measurements (i.e. luminosity and $T_{eff}$), we simply assume that
fractional errors in period are normally distributed with a standard
deviation of 1\%, characteristic of the typical errors in pre-main
sequence stellar period measurements.

For the remaining variables it is possible to ascertain some measure
of the actual errors from the observed data.  Fractional $v \sin i$
errors for the distance sample were calculated using the relationship
between $R$ and $\delta (v \sin i)$ given in
$\S$\ref{testing_uncertainties}.  The resulting error distribution is 
consistent with a normal distribution with $\sigma$=20\%.

Fractional errors in $T_{eff}$ were determined by comparing values of
T$_{eff}$ calculated using individual spectral type-based reddening
corrections or simply adopting the modal reddening for all cluster
members.  The T$_{eff}$ error distribution calculated using this
prescrption is consistent with a normal distribution with $\sigma$ =
10\%.  We expect, however, that calculating T$_{eff}$ values by
adopting the cluster's modal reddening is less accurate than deriving
reddenings from observed spectral types.  Indeed, we note a clear
relationship between the $\delta$T$_{eff}$ value measured for each
star and its spectroscopic reddening estimate. This suggests that the
difference between the two T$_{eff}$ estimates is dominated by the
errors introduced by adopting the modal reddening, and that the
resulting error distribution overestimates the actual errors
associated with the T$_{eff}$ values derived using individual spectral
type-based reddening corrections.


Similarly, the errors in $L$ are estimated from the difference between
luminosities calculated assuming a reddening derived from each star's
spectral type and those calculated assuming an overall reddening for
the cluster.  We find that the luminosity error distribution implied
by these distinct $L$ estimates lies within the bounds of the normal
distribution with $\sigma$=35\% suggested by
\citet{Hartmann_agespreads} as characteristic of luminosity errors in
pre-main sequence stars.


\subsection{Model $\sin i$ Distributions}
\label{construct_model}

Our technique for modeling the distribution of $\sin i$ in NGC 2264
borrows heavily from \citet{Preibisch}.  We define the modeled value
of $\sin i$, $(\sin i)_{m}$ as follows:
\begin{eqnarray}
\label{sinim}
(\sin i)_{m} = \frac{P_{0} (v \sin i)_{0} (T_{eff})_{0}^{2}
    \sqrt{\sigma / \pi}S}{\sqrt{L_{0}}} \\ \times \left[
    \frac{\frac{P}{P_{0}} \frac{(v \sin i)}{(v \sin)_{0}}
    \frac{T_{eff}^{2}}{(T_{eff})_{0}^{2}}}{\frac{\sqrt{L}}{\sqrt{L_{0}}}}
    \right] \nonumber,
\end{eqnarray}
where the subscripted variables represent the actual values
independent of measurement uncertainties, and the non-subscripted
variables represent the observed values including measurement
uncertainties.  Eq. \ref{sinim} allows us to split the dependence of
$\sin i$ into two parts: the actual value of $\sin i$ (the term
outside the brackets) and the contributions of measurement
uncertainties (the term inside the brackets).

\begin{figure}[h]
\centering
\includegraphics[scale = 0.5]{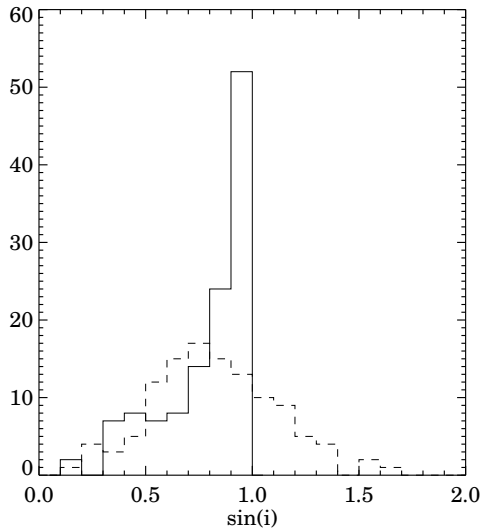}
\caption{Modeled $\sin i$ distribution assuming random axial
  orientations before (solid line) and after (dashed line) taking into 
  account measurement uncertainties.  The measurement
  uncertainties assumed in the generation of the model distribution
  are drawn randomly from the observed uncertainty distributions.  We
  also restrict the model to $v \sin i > 9$ km/s (see
  $\S$\ref{vtrue_distribution}). }
\label{modeledsini}
\end{figure}

The $(\sin i)_{m}$ distribution is generated from Eq. \ref{sinim}
using a Monte Carlo routine.  Each term of the form $X/X_{0}$ (where
$X$ represents any of the variables $P$, $v \sin i$, $T_{eff}$ or $L$)
is calculated by drawing randomly from the appropriate error
distribution for our dataset, as described in
$\S$\ref{error_distribution}.  This process assumes that fractional
errors in $v \sin i$, $T_{eff}$ and $L$ are independent of the values
of these variables.  While this assumption is somewhat questionable,
the deviation from the true error distribution is likely small.  An
example of a $(\sin i)_{m}$ distribution that includes the effects of
observational uncertainties is shown in Fig. \ref{modeledsini}.

\subsubsection{Distribution of $v_{true}$}
\label{vtrue_distribution}

In order to incorporate the effects of the $v \sin i$ cutoff adopted
in $\S$\ref{sampleselection} into our model it is necessary to assume
some prescription for the distribution of the true equatorial
velocities, $v_{true}$.  Once such a prescription has been assumed,
modeled $v \sin i$ values can be calculated, allowing the model sample
to be restricted in the same manner as the observational sample.

\begin{figure}[h]
\centering
\includegraphics[scale = 0.5]{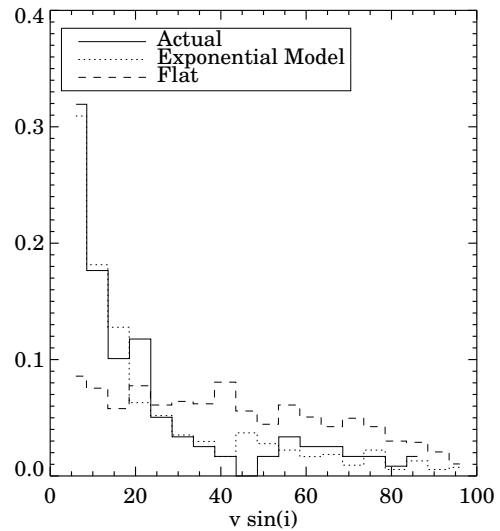}
\caption{Comparison of the $v \sin i$ distributions resulting from
  different assumed $v_{true}$ distributions.  The exponential model
  assumes the probability of a star having a particular $v_{true}$
  goes as an exponentially decaying function, plus a constant offset.
  The flat model assumes that all $v_{true}$ in the range considered
  (0 to 100 km/s) are equally likely.  As can be seen from the figure,
  the exponential model fits well to the data.}
\label{vtrue_distribs}
\vskip 0.1cm
\end{figure}

We tested a variety of $v_{true}$ distributions by combining randomly
sampled $v$ values with random axial orientations and comparing the
resultant set of modeled $v \sin i$ values with the observed $v \sin
i$ values.  The input $v_{true}$ distribution was adjusted until the
sampled population produced a satisfactory match with the observed $v
\sin i$ values, as indicated by a two-sided Kolmogorov-Smirnov (KS)
test.  We find that modeling the $v_{true}$ distribution as an
exponentially decaying function with a constant offset ($P(v_{true})
\propto e^{-\alpha \cdotp v_{true}} + C$) leads to a good match with
the observed $v \sin i$ distribution.  Our best matching $v_{true}$
model has $\alpha = 0.09$ and $C = 0.004$.  A KS test comparing the
resultant modeled $v \sin i$ distribution with the observed
distribution yields, on average, a probability of $\sim 95 \%$ that
our modeled $v \sin i$ distribution comes from the same underlying
distribution as the observed $v \sin i$'s.  A flat distribution of
$v_{true}$, on the other hand, can be rejected with a probability
greater than $99.999 \%$.  See Fig. \ref{vtrue_distribs} for a
comparison of the assumed $v_{true}$ distributions.  This result is in
agreement with that found by \citet{Jeffries}.

\subsubsection{The Binary Correction}

Some unknown fraction, $B$, of the stars included in our catalog are
unresolved binary systems.  For such systems, the value of $L$ that we
calculate will characterize the total system luminosity, not the
luminosity of a single star.  That is, unresolved binaries result in
overestimates of the luminosity of the primary star.  Since the value
of $\sin i$ depends inversely on $\sqrt{ L }$, the presence of
unresolved binaries in our observational sample will cause $\sin i$ to
be systematically underestimated, or conversely, for the value of
$(\sin i)_{m}$ to be systematically overestimated.

To correct $(\sin i)_{m}$ to account for unresolved binaries, we
assume that the masses and luminosities of both the primary and the
secondary can be related through a mass-luminosity relation of the
form
\begin{eqnarray}
L = k M^{a},
\end{eqnarray}
where $k$ and $a$ are constants.  Thus, we can express the total luminosity
of the primary and secondary, $L_{T}$, as
\begin{eqnarray}
\label{binfraceqn}
L_{T} = L_{P} \left( 1 + q^{a} \right),
\end{eqnarray}
where $L_{P}$ is the luminosity of the primary and $q$ is the mass
ratio of the secondary to the primary, $q = M_{s} / M_{p} \leq 1$.
Assuming that the temperature derived for the binary system
corresponds to the temperature of the primary, then the value of $(
\sin i )_m$ calculated for the primary star in a binary system can be
corrected by simply dividing by a correction factor of $\sqrt{1 +
q^{a} }$.  Since we do not know the exact values of $q$, we simply
assume that $q$ is uniformly distributed between $0$ and $1$.  While
the true distribution of $q$ is almost certainly not uniform, the
deviation from a uniform distribution is likely small enough to not
significantly impact our results.

To determine the value of $a$ in Eq. \ref{binfraceqn} we fit to the
mass and luminosity models derived for pre-main sequence stars between
the ages of 1-10 Myr by \citet{Baraffe}.  From this fit we derive a
value of $a=1.5$.

The binary fraction for NGC 2264 is poorly constrained.  The
multiplicity of nearby main-sequence field stars appears to be mass
dependent, ranging from $\sim$50\% for G stars \citep{Duquennoy1991}
to $\sim$30\% for M stars \citep{Fischer1992,Reid1997,Delfosse2004}.
The multiplicity for regions of isolated star formation, such as
Taurus, has been found to be considerably higher than the field
population, perhaps larger than 80\% \citep{Leinert}.  Clustered
regions such as the ONC and NGC 2264, however, show no such excess
\citep{Kohler2006}, suggesting initial conditions or dynamical
evolution at early ages have an important effect of stellar
multiplicity in the pre-main sequence phase.  As NGC 2264 is closer in
character to the ONC than Taurus, we assume a value of $B = 0.50$ as
the preferred binary fraction for our model, and investigate values as
low as $B = 0.0$ and as high as $B = 0.75$.

\subsection{Comparing Model $\sin i$ Distributions with Observations}
\label{kolmogorov}

\subsubsection{Measuring a best fit distance}

\begin{figure}[h]
\centering
\includegraphics[scale = 0.5]{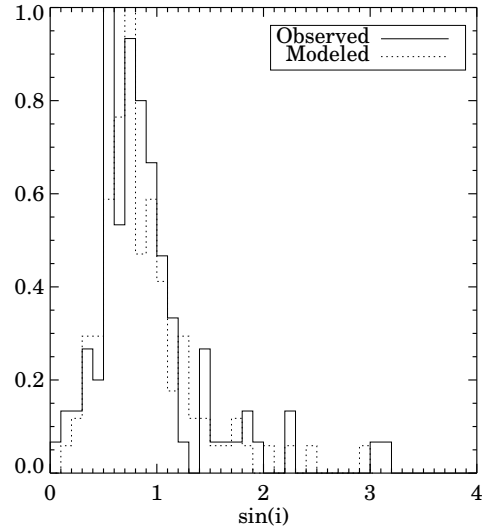}
\caption{Comparison of the observed distribution of $\sin i$ assuming
  a distance to NGC 2264 of 913 pc with the modeled distribution.}
\label{sinicompare}
\end{figure}

\begin{figure}[h]
\centering
\includegraphics[scale = 0.5]{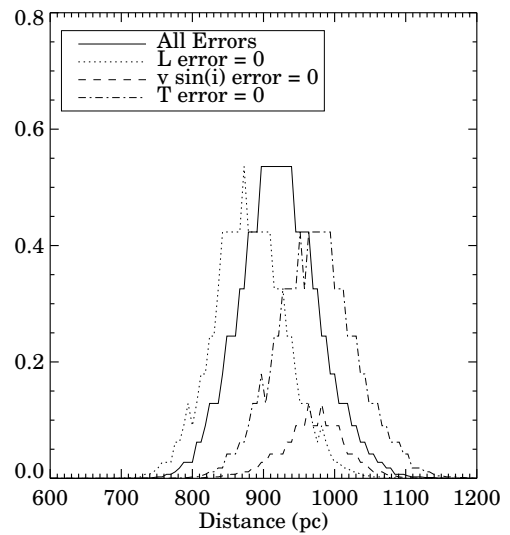}
\caption{Effect of various assumed error models on the $P_{KS}$
  vs. distance curve.  Vertical axis represents the probability that
  the observed $\sin i$ distribution at a particular distance was
  drawn from the same distribution as the modeled $\sin i$ distribution.}
\label{ks_errs}
\end{figure}

We compare the modeled distribution of $\sin i$ to the observed
distribution using two-sided KS tests over a range of assumed
distances.  The KS tests were repeated 100 times at distances of 600
pc to 1100 pc in steps of 5 pc.  The distance with the highest median
KS probability, $P_{KS}$, is the best fit distance.
Fig. \ref{ks_errs} shows the $P_{KS}$ vs. distance curve for the most
likely set of adopted model parameters (solid line: $B = 0.5$, all
errors drawn from the observed error distributions).  We find the best
fit distance to be $913$ pc.  At this distance, the median KS
probability is $\sim 0.5$.  The observed and modeled distributions
corresponding to this best fit distance are shown in
Fig. \ref{sinicompare}.

\subsubsection{Measuring statistical errors with a bootstrap analysis}

We have performed a bootstrap analysis to obtain a rigorous estimate
of the statistical uncertainties associated with our measurement of
the distance to NGC 2264.  In the bootstrap procedure, many artificial
$\sin i$ datasets are generated by selecting with replacement from the
actual set of observed $\sin i$ values.  Each of these synthetic data
sets are analyzed using the method described in $\S$\ref{kolmogorov}
to determine a best-fit distance to the cluster.  The resultant
distribution of derived distances provides an estimate of the true
distance, and the width of the distribution provides an estimate for
the uncertainty in our derived distance.  By using replacement, the
bootstrap procedure tests the uncertainty in our distance estimate due
to sampling effects, but does not account for any selection effects or
biases that would introduce systematic differences between the true
and observed $\sin i$ distributions.

\begin{figure}[h]
\centering
\includegraphics[scale = 0.5]{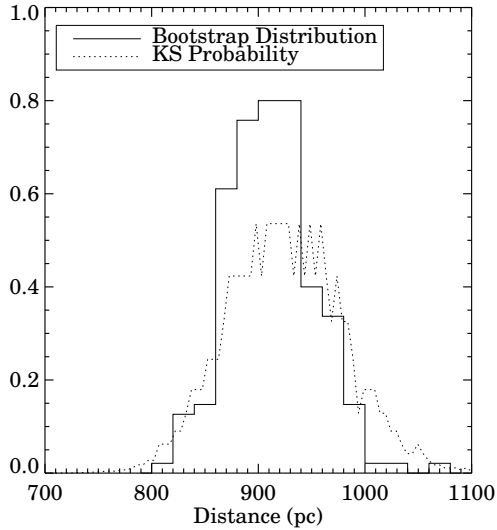}
\caption{The distribution of bootstrapped best-fit distances is
  overlaid on a plot of the $P_{KS}$ vs. distance curve for the
  observed data.  A binary fraction of $B = 0.5$ is assumed, and
  errors are drawn randomly from the observed error distributions.}
\label{bootstrapfinal}
\end{figure}

For the bootstrap analysis, we generated 200 sets of $\sin i$ values
of equal size to the observed sample (i.e. 97 stars) by randomly
selecting values with replacement from the list of observed $\sin i$
values.  For each of the 200 sets, a best fit distance was derived
using the KS procedure described above.  The resultant distribution of
distance estimates is shown in Fig. \ref{bootstrapfinal}.  We derive a
1-$\sigma$ confidence range for our distance estimate from the width
of the velocity region that encloses $67 \%$ of the best-fit distances
measured with the bootstrap procedure.  Combining this uncertainty
estimate with the best fit distance measured in $\S$\ref{kolmogorov}
yields a distance estimate to NGC 2264 of $913 \pm 40$ pc.

\subsubsection{Investigating systematic effects}

Our distance measurement includes statistical
uncertainties due to sampling effects, as calculated above, but 
also potential systematic errors due to the assumptions that 
underlie our models.  Factors that could introduce systematic errors
into our analysis include: the error prescriptions and binary fraction 
adopted in our calculation of $(\sin i)_{m}$; the $v \sin i$ cutoff 
we imposed on our modeled and observed $\sin i$ distributions; biases 
in the stellar properties derived for stars as a function of their 
evolutionary state; and finally, the underlying assumption of isotropically 
distributed rotation axes.  We consider in turn the potential impact of each 
of these effects on our analysis.

The impact of different sources of observational error on the distance
determination can be seen in Fig. \ref{ks_errs}, where we compare the
$P_{KS}$ vs. distance relations produced by comparing our observed
sample to models that neglect various components of the $\sin i$
error budget.  The solid line represents the case where all of the
errors are chosen randomly from the actual error distributions, and
the remaining curves represent cases where the errors in a particular
variable have been set to 0.  It is clear from Fig. \ref{ks_errs} that
the assumed errors have a non-negligible impact on both the best-fit
distance as well as the height of the $P_{KS}$ curve.  For instance,
negating the luminosity errors has the effect of reducing the best-fit
distance by 4 \%, while eliminating the $T_{eff}$ errors
increases the best-fit distance by roughly 6 \%.  Eliminating the $v
\sin i$ errors not only increases the best fit 
distance by 4 \%, but also strongly reduces the peak $P_{KS}$, 
suggesting that the errors in $v \sin i$ contribute
significantly to the shape of the $\sin i$ distribution.

To the extent that the error distributions we adopt in our model (see
$\S$\ref{error_distribution}) do not reflect the true error
distributions of our data, our distance estimate will be skewed.  As
shown above, the error distributions we do adopt are influencing our
derived distance at the $\sim$5\% level, suggesting any systematic
error in our derived distance due to adopting improper error
distributions would likely be $\sim$5\% as well.  Our error
distributions, however, are consistent with those estimated by other
authors for the same parameters, and the error distributions push the
derived distance in different directions, such that multiple
systematic errors should offset one another to some degree.  Perhaps
most worrisome is the asymmetric shape of the luminosity error
distribution, but lacking a more robust means of characterizing the
luminosity errors, we are unable to remove this potential systematic
effect from our analysis.

\begin{figure}[h]
\centering
\includegraphics[scale = 0.5]{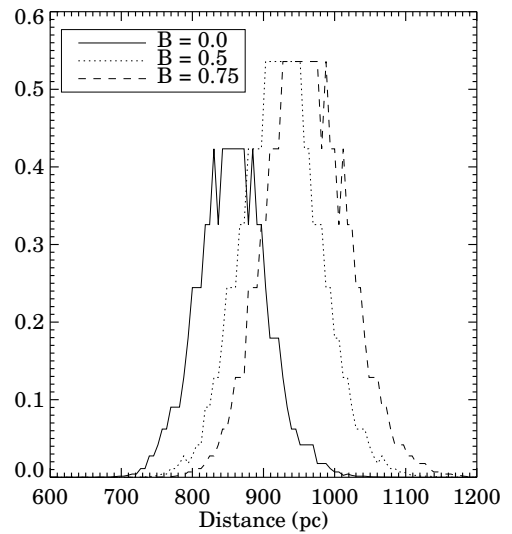}
\caption{Effect of various assumed binary fractions, $B$, on the
  $P_{KS}$ vs. distance curve.  Vertical axis represents the
  probability that the observed $\sin i$ distribution at a particular
  distance was drawn from the same distribution as the modeled $\sin
  i$ distribution.}
\label{ks_binary}
\end{figure}

Fig. \ref{ks_binary} shows the effect of the assumed binary fraction
on the best fit distance.  As the binary fraction is increased, the
average luminosity of the modeled stars is increased, causing the
$\sin i$ distribution to move to lower $\sin i$ as per
Eq. \ref{sinim}.  In order for the observed $\sin i$ distribution to
remain well matched to the modeled distribution, the assumed distance
must therefore increase according to Eq. \ref{siniL}.  Thus, we see
that as the modeled binary fraction changes from $B = 0.0$ to $B = 0.75$, the
best-fit distance goes from $\sim 860 $ pc to $\sim 970$ pc, a change
of roughly 13\%.  This is consistent with the results 
of \citet{Jeffries}, who found in his ONC study that increasing the binary
fraction by 0.2 led to a 4\% increase in his modeled best fit distance. 

Biases in the parameters derived for stars in different evolutionary
states could also the results of our analysis.  As noted by
\citet{Jeffries}, it may be more difficult to derive accurate
temperatures, luminosities, and, thus, radii for classical T Tauri
stars than more evolved weak T Tauri stars; classical T Tauri stars
typically possess larger extinctions, are irregularly variable, and
have significant contributions to their total luminosity from
accretion and re-radiation from their circumstellar disk, all of which
can complicate the derivation of their intrinsic stellar properties.
Following \citet{Lada2006}, we used an IRAC [3.6 - 8] color cut to
identify classical T Tauri stars in our sample; when these CTTSs are
excluded from our sample, the best fit distance to NGC 2264 increases
to $\sim$950 pc.  This effect is contrary to that seen by
\citet{Jeffries}, who found a decreased best fit distance to the ONC
once CTTSs were excluded from his sample.  Given these contradictory
results, and that the effect of removing CTTSs from our sample is
comparable to the other systematic effects probed here, we have chosen
not to exclude CTTSs from our main analysis.

Our technique assumes that all of the stars in NGC 2264
are at the same distance from Earth.  In reality, the cluster has some
line of sight depth, predicted to be on the order of 28 pc \citep{Dahm}.
If we assume that the known members of NGC 2264 are biased towards low
extinctions, this could cause the distance we derive to the cluster to be 
slightly smaller than the true geometric center of the cluster.  
The distance we derive here, however, would properly describe the distance
of the population of currently known members; we do
not consider this to be a systematic error in our analysis, but rather
a nuance that should inform the interpretation of our results.

Finally, the fundamental assumption of random axial orientations is itself
somewhat questionable since stellar clusters like NGC 2264 are
predicted to have collapsed from single cloud complexes.  We might
expect, then, for there to be a preferred orientation of stars in the
cluster resulting from the conservation of the cloud's initial angular
momentum, or possibly from the presence of large scale magnetic
fields.  Previous applications of the $\sin i$ distribution technique
have, however, produced results that are in agreement with precise
parallax measurements \citep{Preibisch, Jeffries}.  This agreement
provides evidence in support of the random axial orientation
assumption, but it has not been directly confirmed by observations.  

The discussion above has revealed a number of potential
systematic effects in our analysis.  The five potential systematics 
most amenable to direct investigation (the inclusion of CTTSs in 
our primary sample; the adopted luminosity, temperature, and 
$v \sin i$ error distributions; and the assumed binary 
fraction) all influence the derived best fit distance at the $\sim$ 5\% 
level.  We therefore combine these individual uncertanties in quadrature 
to characterize the potential error in our best fit distance
due to systematic effects, producing a estimate of our total systematic
uncertainties of $\pm$ 12 \%, or 110 pc.

\begin{deluxetable*}{cccc}[!h]
\tablewidth{0pt}
\tabletypesize{2pt}
\tablecaption{Comparison of Previous Distances Estimates for NGC 2264 
\label{tab:prevdist}}
\tablehead{
   \colhead{Authors} &
   \colhead{Distance (pc)} &
   \colhead{Method} &
   \colhead{Number of Stars}}
\startdata
\citet{Perez} & $950 \pm 75$ & Cluster fitting & $21$ \\
\citet{Neri} & $910 \pm 50$ & Cluster fitting & $\sim 50$ \\
\citet{Sung} & $760 \pm 85$ & Cluster fitting & 13 \\ 
This work & $913 \pm 40$ & $\sin i$ statistical technique & 97 \\
\enddata
\end{deluxetable*}

\subsection{The derived distance and age of NGC 2264}
\label{discussion}

We have calculated distance-dependent $\sin i$ values for a sample of
97 pre-main sequence stars in the open cluster NGC 2264. By comparing
the observed $\sin i$ distribution to a modeled distribution assuming
random axial orientations, we derive a distance of 913 pc to NGC 2264;
quantitative tests of our analysis reveal sampling and systematic
errors of 40 and 110 pc, respectively.  Our distance estimate does not
rely on evolutionary models to any significant degree.\footnote{The
one exception is the use of the pre-main sequence mass-luminosity
relation in the derivation of the binary correction factor.  This
factor had a relatively small effect on our calculated distance,
however, and is also well constrained by observations.  Thus, our
derived distance is almost entirely model-independent.}

Our distance estimate is significantly higher than a number of
previously determined distances, particularly the widely cited value
of 760 pc found by \citet{Sung}.  In general, though, our estimate
falls within the typical range of calculated distances (730 pc
to 950 pc) for NGC 2264.  Table \ref{tab:prevdist} provides a
comparison of our result to previous distance estimates.
A distance to NGC 2264 of 913 pc represents an increase of
approximately 20\% compared to the widely excepted value of 760 pc, though 
the two results are formally consistent within the sum of the statistical and
systematic error bars.  

The mean age of NGC 2264 is commonly cited as $\sim$ 3 Myrs, though
there is evidence for a considerable age spread within the cluster
\citep{Dahm}.  The luminosities of pre-main sequence stars are often
compared with predictions of theoretical pre-main sequence models to
infer the age of their parent cluster; as luminosity declines through
the pre-main sequence phase, the larger luminosities produced by
assuming a greater distance to the cluster will produce a younger
inferred age for the cluster.  Increasing the assumed distance to NGC
2264 from 760 to 910 pc changes the distance modulus by 0.4 mag.  The
corresponding 0.4 mag brightening of the stars produces a shift in the
age of the cluster. We have produced a crude estimate of the revised
age of the cluster by determining the age at which a 1 M$_{\odot}$
star's H band magnitude is 0.4 mag brighter than at 3 Myrs: according
to the pre-main sequence models calculated by \citet{Baraffe}, the
distance we derive here implies a revised age for NGC 2264 of
$\sim$1.5 Myrs.  The detailed analysis of cluster members required for
robust estimates of the age and properties of NGC 2264 in light of our
new derived distance, however, is beyond the scope of this work.

\section{Conclusions}
We determined the distance to the open cluster NGC 2264 using a
statistical analysis of cluster member inclinations.  We derived
distance-dependent values of $\sin i$ (where $i$ is the inclination
angle) for 97 stars in NGC 2264 from measured rotation periods,
luminosities, effective temperatures, and projected equatorial
rotation velocities, $v \sin i$, of these stars.  We measured 96
of the $v \sin i$ values in our sample by analyzing high-resolution
spectra with a cross-correlation technique.  We modeled the observed
distribution of $\sin i$ for the cluster by assuming that member stars
have random axial orientations and by adopting prescriptions for the
measurement errors in our sample.  By adjusting the distance assumed
in the observed $\sin i$ distribution until it matches the modeled
distribution, we obtained a best-fit distance for the cluster.  We find
the data to be consistent with a distance to NGC 2264 of
$913$ pc.  Quantitative tests of our analysis reveals uncertainties of 40 and
110 pc due to sampling and systematic effects, respectively.  This
distance estimate suggests a revised age for the cluster of $\sim$1.5 Myrs,
although more detailed investigations of the full cluster membership are
required to draw strong conclusions.

\acknowledgements The authors wish to thank Steve Strom for a prompt and
helpful referee report that improved the analysis presented here, and 
Russel White and Jeff Burchfield for providing IDL code that formed 
the basis of our cross correlation pipeline.  EJB acknowledges the 
support of the SAO Summer Intern Program, made possible by a grant 
from the NSF.  NASA support was provided to K. Covey for this work 
through the Spitzer Space Telescope Fellowship Program, through a 
contract issued by the Jet Propulsion Laboratory, California 
Institute of Technology under a contract with NASA.

\LongTables
\begin{deluxetable}{cccccccccc}[!htp]
\tablewidth{0pt}
\tabletypesize{\scriptsize}
\tablecaption{Data used to create $\sin i$ distribution 
\label {tab:sinidata}}
\tablehead{
   \colhead{RA} &
   \colhead{DEC} &
   \colhead{$\log L_{bol}$ } &
   \colhead{$\log T_{eff}$} &
   \colhead{$L_{bol}$ and $T_{eff}$} &
   \colhead{Period} &
   \colhead{$v \sin i$} &
   \colhead{$v \sin i$ Error} \\
   \colhead{(deg)} &
   \colhead{(deg)} &
   \colhead{(ergs/sec) } &
   \colhead{(K)} &
   \colhead{Source} &
   \colhead{(days)} &
   \colhead{(km/s)} &
   \colhead{(km/s)}}
\startdata
     99.8594131     &        9.6863384     &    33.590     &   3.630     &       S     &    5.51     &    16.5     &    1.51     \\  
     99.8765793     &        9.5604086     &    33.530     &   3.640     &       S     &    5.49     &    57.5     &    6.39     \\  
     99.9138336     &        9.9332304     &    33.420     &   3.590     &       S     &    0.86     &    66.9     &    5.33     \\  
     99.9232178     &        9.5779305     &    34.180     &   3.760     &       S     &    3.61     &    24.6     &    2.18     \\  
     99.9446716     &        9.6816502     &    34.160     &   3.740     &       S     &    3.84     &    24.6     &    2.43     \\  
     99.9566879     &        9.5561504     &    33.470     &   3.650     &       S     &    6.53     &    10.5     &    4.08     \\  
    100.0046692     &        9.5926476     &    33.000     &   3.540     &       P     &    9.04     &    21.6     &   13.24     \\  
    100.0111237     &        9.5900726     &    33.100     &   3.540     &       P     &    4.65     &    14.1     &   28.76     \\  
    100.0250320     &        9.8285141     &    33.160     &   3.560     &       S     &    8.12     &    13.9     &    4.70     \\  
    100.0429306     &        9.6486139     &    33.810     &   3.630     &       S     &    3.83     &    33.9     &    1.25     \\  
    100.0453644     &        9.6686916     &    32.800     &   3.540     &       S     &   11.73     &    11.1     &   25.58     \\  
    100.0804520     &        9.8083334     &    33.540     &   3.590     &       S     &    1.03     &    62.1     &    3.60     \\  
    100.0844803     &        9.9350996     &    33.600     &   3.540     &       S     &    3.87     &    25.7     &   20.12     \\  
    100.0988846     &        9.9232969     &    33.200     &   3.530     &       P     &    4.57     &    15.3     &   27.21     \\  
    100.1061554     &        9.8072138     &    33.620     &   3.640     &       S     &    3.14     &    25.0     &    2.20     \\  
    100.1191101     &        9.5965662     &    33.530     &   3.590     &       S     &    4.57     &    15.8     &    1.97     \\  
    100.1206131     &        9.7047615     &    33.480     &   3.620     &       S     &    7.22     &    16.0     &    1.17     \\  
    100.1275787     &        9.7696247     &    33.600     &   3.670     &       S     &    7.23     &    12.5     &    1.07     \\  
    100.1285858     &        9.5779390     &    33.810     &   3.680     &       S     &   12.09     &    10.5     &    8.96     \\  
    100.1318436     &        9.8064947     &    33.440     &   3.640     &       S     &    4.17     &    23.5     &    1.75     \\  
    100.1366730     &        9.8581495     &    33.580     &   3.590     &       S     &    3.46     &    30.1     &    9.57     \\  
    100.1436234     &        9.5884171     &    33.240     &   3.550     &       S     &    3.88     &    85.9     &   15.74     \\  
    100.1521683     &        9.8460083     &    33.370     &   3.550     &       S     &    7.79     &     9.6     &    5.37     \\  
    100.1526184     &        9.8063803     &    33.470     &   3.630     &       S     &   16.49     &    11.4     &    3.64     \\  
    100.1528091     &        9.7895918     &    35.730     &   3.930     &       S     &    4.12     &    87.9     &   33.18     \\  
    100.1550217     &        9.5194111     &    33.220     &   3.550     &       P     &    1.15     &    15.7     &    5.25     \\  
    100.1668930     &        9.5841417     &    33.680     &   3.660     &       S     &    4.50     &    13.1     &    0.95     \\  
    100.1723404     &        9.9038496     &    34.240     &   3.760     &       S     &    3.42     &    41.2     &    2.19     \\  
    100.1806259     &        9.8498755     &    33.580     &   3.630     &       S     &    9.04     &    11.2     &    1.01     \\  
    100.1838608     &        9.3987112     &    33.490     &   3.690     &       S     &    2.26     &    37.6     &    6.33     \\  
    100.1868820     &        9.9622917     &    33.790     &   3.640     &       S     &   16.05     &     9.1     &    0.84     \\  
    100.1876907     &        9.7616167     &    33.480     &   3.630     &       S     &    4.61     &    21.4     &    2.84     \\  
    100.1920166     &        9.8214893     &    34.790     &   3.730     &       S     &    0.74     &    75.9     &    1.51     \\  
    100.2003937     &        9.8942642     &    33.330     &   3.650     &       S     &    5.43     &    12.7     &    1.46     \\  
    100.2011337     &        9.6107359     &    33.360     &   3.550     &       S     &    1.67     &    52.0     &    9.42     \\  
    100.2035065     &        9.7237997     &    33.110     &   3.550     &       P     &    9.04     &    15.1     &    8.61     \\  
    100.2144318     &        9.6206837     &    33.200     &   3.580     &       S     &    8.94     &     9.1     &    2.53     \\  
    100.2194901     &        9.7391720     &    33.220     &   3.550     &       S     &    5.41     &    14.4     &    3.27     \\  
    100.2234802     &        9.5568609     &    33.620     &   3.720     &       S     &    2.38     &    27.6     &    1.23     \\  
    100.2260971     &        9.8223219     &    33.240     &   3.550     &       S     &    9.80     &    16.2     &    5.51     \\  
    100.2481079     &        9.5863609     &    34.180     &   3.630     &       S     &    3.35     &    43.3     &    2.07     \\  
    100.2500000     &        9.4805641     &    33.560     &   3.630     &       S     &    4.18     &    22.9     &    2.29     \\  
    100.2521362     &        9.4877644     &    33.820     &   3.680     &       S     &    5.22     &    40.9     &    2.35     \\  
    100.2532425     &        9.8562031     &    34.040     &   3.710     &       S     &    4.41     &    20.2     &    1.71     \\  
    100.2607727     &        9.5869751     &    33.490     &   3.640     &       S     &    4.24     &    17.7     &    1.42     \\  
    100.2642822     &        9.5013723     &    33.380     &   3.530     &       P     &    1.31     &    64.4     &   13.03     \\  
    100.2645721     &        9.5217781     &    34.310     &   3.760     &       S     &    2.18     &    61.0     &    2.64     \\  
    100.2648849     &       10.0098276     &    33.220     &   3.570     &       S     &   11.20     &    12.6     &    2.56     \\  
    100.2650299     &        9.5080585     &    33.320     &   3.510     &       P     &    9.71     &    26.1     &   71.63     \\  
    100.2668304     &        9.8191080     &    33.810     &   3.670     &       S     &   12.43     &    13.8     &    1.12     \\  
    100.2680588     &        9.8061390     &    33.780     &   3.700     &       S     &    1.32     &    20.3     &    0.96     \\  
    100.2683716     &        9.8639193     &    34.460     &   3.720     &       S     &    3.75     &    33.7     &    1.22     \\  
    100.2707062     &        9.8461361     &    33.910     &   3.700     &       S     &    3.70     &    33.0     &    1.78     \\  
    100.2712402     &        9.8133221     &    33.430     &   3.570     &       S     &    3.57     &    16.1     &    2.80     \\  
    100.2712479     &        9.8623857     &    33.640     &   3.670     &       S     &    9.88     &    11.1     &    0.98     \\  
    100.2723541     &        9.5537281     &    33.610     &   3.640     &       S     &    1.20     &    27.4     &    2.08     \\  
    100.2740784     &        9.8048582     &    34.030     &   3.680     &       S     &    8.46     &    13.4     &    2.94     \\  
    100.2742157     &        9.8799639     &    33.510     &   3.620     &       S     &    4.21     &    22.3     &    1.95     \\  
    100.2758408     &        9.6063833     &    34.120     &   3.720     &       S     &    3.00     &     9.7     &    1.54     \\  
    100.2787323     &        9.4900112     &    33.240     &   3.600     &       S     &    3.14     &    10.2     &    3.91     \\  
    100.2798233     &        9.4633245     &    33.970     &   3.660     &       S     &    9.61     &    13.4     &    0.90     \\  
    100.2802658     &        9.9753218     &    33.690     &   3.650     &       S     &    2.59     &    34.5     &    2.79     \\  
    100.2823486     &        9.6874838     &    33.490     &   3.570     &       S     &    1.97     &    20.8     &    1.87     \\  
    100.2833862     &        9.5112028     &    33.450     &   3.680     &       S     &    3.88     &    20.5     &    2.11     \\  
    100.2868195     &        9.3952942     &    33.790     &   3.600     &       S     &    1.83     &    67.9     &    2.41     \\  
    100.2873383     &        9.5627832     &    33.450     &   3.530     &       S     &    0.80     &    76.9     &   92.72     \\  
    100.2895279     &        9.8639059     &    33.680     &   3.550     &       S     &    5.92     &    67.3     &    5.20     \\  
    100.2958221     &        9.5988054     &    33.560     &   3.600     &       S     &   11.08     &    17.5     &   28.98     \\  
    100.3050003     &        9.4362030     &    34.140     &   3.670     &       S     &    1.76     &    56.5     &    2.94     \\  
    100.3102951     &        9.5559502     &    33.590     &   3.600     &       S     &   11.07     &     9.4     &    0.84     \\  
    100.3156281     &        9.4380083     &    34.190     &   3.660     &       S     &    6.23     &    28.8     &    1.63     \\  
    100.3199310     &        9.4583778     &    33.580     &   3.660     &       S     &    5.20     &    13.3     &    2.09     \\  
    100.3237991     &        9.4906082     &    33.750     &   3.560     &       S     &    2.43     &    60.9     &    4.54     \\  
    100.3246765     &        9.4836388     &    33.300     &   3.530     &       S     &    2.92     &    71.8     &   11.75     \\  
    100.3246994     &        9.5602837     &    33.600     &   3.610     &       S     &    6.51     &    10.2     &    3.83     \\  
    100.3265533     &        9.6614218     &    33.300     &   3.600     &       S     &   11.32     &    54.5     &   15.99     \\  
    100.3317947     &        9.5289888     &    33.940     &   3.650     &       S     &    2.50     &    36.1     &    2.07     \\  
    100.3356018     &        9.7598753     &    33.740     &   3.710     &       S     &    1.30     &    24.2     &    4.64     \\  
    100.3417587     &        9.7202024     &    33.540     &   3.680     &       S     &    6.51     &    10.0     &    2.67     \\  
    100.3460236     &        9.7240610     &    33.170     &   3.540     &       S     &    0.94     &    62.5     &   19.33     \\  
    100.3625031     &        9.5036526     &    33.220     &   3.570     &       S     &    5.08     &    16.8     &    5.54     \\  
    100.3631363     &        9.5850334     &    33.840     &   3.700     &       S     &    1.54     &    86.8     &    3.38     \\  
    100.3699265     &        9.6441221     &    33.920     &   3.700     &       P     &    5.89     &    22.1     &    0.68     \\  
    100.3816910     &        9.8091164     &    33.860     &   3.700     &       S     &    5.06     &    21.6     &    1.73     \\  
    100.3833160     &       10.0068026     &    33.480     &   3.650     &       S     &    4.74     &    16.7     &    2.25     \\  
    100.4053726     &        9.7518587     &    33.700     &   3.680     &       S     &    4.58     &    11.5     &    1.14     \\  
    100.4280701     &        9.7157307     &    33.640     &   3.620     &       S     &    3.93     &    18.8     &    1.61     \\  
    100.4286575     &        9.4189997     &    33.130     &   3.590     &       P     &    4.74     &    10.3     &    3.97     \\  
    100.4435120     &        9.7185698     &    33.390     &   3.640     &       S     &    4.30     &    14.9     &    4.71     \\  
    100.4502716     &        9.7120361     &    34.230     &   3.750     &       S     &    1.19     &    91.0     &    2.78     \\  
    100.4542313     &        9.6850386     &    33.510     &   3.580     &       S     &    4.48     &    15.6     &    1.89     \\  
    100.4584656     &        9.4922886     &    33.510     &   3.640     &       S     &    3.27     &    29.1     &    5.66     \\  
    100.4644318     &        9.8951693     &    33.870     &   3.720     &       S     &    3.70     &    23.3     &    1.01     \\  
    100.4644928     &        9.7360191     &    33.200     &   3.570     &       S     &    0.68     &     9.5     &    2.99     \\  
    100.4710388     &        9.9674664     &    33.780     &   3.630     &       S     &    0.93     &    80.8     &    1.89     \\  
    100.4714890     &        9.8465023     &    33.440     &   3.630     &       S     &    5.77     &     9.2     &    2.35     \\  
    100.4918365     &        9.7184000     &    33.560     &   3.640     &       S     &    2.11     &    21.6     &    2.49     \\  
\enddata
\end{deluxetable}

\end{document}